\documentclass[preprint]{aastex}
\usepackage[english]{babel}
\usepackage{graphicx}
\slugcomment{Accepted for publication in ApJL}
\shorttitle{The discovery of CR trapped antiprotons}
\shortauthors{Adriani {\em et al.}}
\usepackage{hyperref}
\begin{document}

\title{The discovery of geomagnetically trapped cosmic ray antiprotons}

\author{O. Adriani$^{1,2}$, G. C. Barbarino$^{3,4}$, G. A. Bazilevskaya$^5$, R. Bellotti$^{6,7}$, M. Boezio$^8$, \\
E. A. Bogomolov$^9$, M. Bongi$^{2}$, V. Bonvicini$^{8}$, S. Borisov$^{10,11,12}$, S. Bottai$^{2}$, A. Bruno$^{6,7,}$\altaffilmark{18}, \\
F. Cafagna$^{6}$, D. Campana$^{4}$, R. Carbone$^{4,11}$, P. Carlson$^{13}$, M. Casolino$^{10}$, G. Castellini$^{14}$, \\
L. Consiglio$^{4}$, M. P. De Pascale$^{10,11}$, C. De Santis$^{10,11}$, N. De Simone$^{10,11}$, V. Di Felice$^{10}$, \\
A. M. Galper$^{12}$, W. Gillard$^{13}$, L. Grishantseva$^{12}$, G. Jerse$^{8,15}$, A. V. Karelin$^{12}$, \\
M. D. Kheymits$^{12}$, S. V. Koldashov$^{12}$, S. Y. Krutkov$^{9}$, A. N. Kvashnin$^{5}$, A. Leonov$^{12}$, \\
V. Malakhov$^{12}$, L. Marcelli$^{10}$, A. G. Mayorov$^{12}$, W. Menn$^{16}$, V. V. Mikhailov$^{12}$, \\
E. Mocchiutti$^{8}$, A. Monaco$^{6,7}$, N. Mori$^{1,2}$, N. Nikonov$^{9,10,11}$, G. Osteria$^{4}$, F. Palma$^{10,11}$, \\
P. Papini$^{2}$, M. Pearce$^{13}$, P. Picozza$^{10,11}$, C. Pizzolotto$^{8}$, M. Ricci$^{17}$, S. B. Ricciarini$^{2}$, \\
L. Rossetto$^{13}$, R. Sarkar$^{8}$, M. Simon$^{16}$, R. Sparvoli$^{10,11}$, P. Spillantini$^{1,2}$, Y. I. Stozhkov$^{5}$, \\
A. Vacchi$^{8}$, E. Vannuccini$^{2}$, G. Vasilyev$^{9}$, S. A. Voronov$^{12}$, Y. T. Yurkin$^{12}$, J. Wu$^{13,}$\altaffilmark{19}, \\
G. Zampa$^{8}$, N. Zampa$^{8}$, and V. G. Zverev$^{12}$}

\affil{$^{1}$University of Florence, Department of Physics, I-50019 Sesto Fiorentino, Florence, Italy}
\affil{$^{2}$INFN, Sezione di Florence, I-50019 Sesto Fiorentino, Florence, Italy}
\affil{$^{3}$University of Naples ''Federico II'', Department of Physics, I-80126 Naples, Italy}
\affil{$^{4}$INFN, Sezione di Naples, I-80126 Naples, Italy}
\affil{$^{5}$Lebedev Physical Institute, RU-119991, Moscow, Russia}
\affil{$^{6}$University of Bari, Department of Physics, I-70126 Bari, Italy}
\affil{$^{7}$INFN, Sezione di Bari, I-70126 Bari, Italy}
\affil{$^{8}$INFN, Sezione di Trieste, I-34149 Trieste, Italy}
\affil{$^{9}$Ioffe Physical Technical Institute, RU-194021 St. Petersburg, Russia}
\affil{$^{10}$INFN, Sezione di Rome ''Tor Vergata'', I-00133 Rome, Italy}
\affil{$^{11}$University of Rome ''Tor Vergata'', Department of Physics, I-00133 Rome, Italy}
\affil{$^{12}$NRNU MEPhI, RU-115409 Moscow, Russia}
\affil{$^{13}$KTH, Department of Physics, and the Oskar Klein Centre for Cosmoparticle Physics, AlbaNova University Centre, SE-10691 Stockholm, Sweden}
\affil{$^{14}$IFAC, I-50019 Sesto Fiorentino, Florence, Italy}
\affil{$^{15}$University of Trieste, Department of Physics, I-34147 Trieste, Italy}
\affil{$^{16}$Universitat Siegen, Department of Physics, D-57068 Siegen, Germany}
\affil{$^{17}$INFN, Laboratori Nazionali di Frascati, Via Enrico Fermi 40, I-00044 Frascati, Italy}

\altaffiltext{18}{Author to whom any correspondence should be addressed: alessandro.bruno@ba.infn.it.}
\altaffiltext{19}{On leave from School of Mathematics and Physics, China University of Geosciences, CN-430074 Wuhan, China.}

\begin{abstract}
The existence of a significant flux of antiprotons confined to Earth's magnetosphere has been considered in
several theoretical works. These antiparticles are produced in nuclear interactions of energetic cosmic rays with
the terrestrial atmosphere and accumulate in the geomagnetic field at altitudes of several hundred kilometers. A
contribution from the decay of albedo antineutrons has been hypothesized in analogy to proton production by
neutron decay, which constitutes the main source of trapped protons at energies above some tens of MeV. This
Letter reports the discovery of an antiproton radiation belt around the Earth. The trapped antiproton energy spectrum
in the South Atlantic Anomaly (SAA) region has been measured by the PAMELA experiment for the kinetic energy
range 60--750 MeV. A measurement of the atmospheric sub-cutoff antiproton spectrum outside the radiation belts
is also reported. PAMELA data show that the magnetospheric antiproton flux in the SAA exceeds the cosmic-ray
antiproton flux by three orders of magnitude at the present solar minimum, and exceeds the sub-cutoff antiproton
flux outside radiation belts by four orders of magnitude, constituting the most abundant source of antiprotons near
the Earth.
\end{abstract}

\section{Introduction}
The PAMELA collaboration has recently reported the cosmic ray (CR) antiproton spectrum and antiproton-to-proton ratio
measurements in the kinetic energy range 60 MeV--180 GeV \citep{PRL,PRL_BIS}. These data significantly improve
those from previous experiments thanks to the high statistical significance and wide energy interval. The results agree with
models of purely secondary production where antiprotons are produced through interactions of CRs with the interstellar
medium.

Antiprotons are also created in pair production processes in reactions of energetic CRs with Earth's exosphere. Some of the antiparticles produced in the innermost region of the magnetosphere are captured by the geomagnetic field allowing the
formation of an antiproton radiation belt around the Earth. The particles accumulate until they are removed due to annihilation
or ionization losses. The trapped particles are characterized by a narrow pitch angle\footnote{The pitch angle is the angle between particle velocity vector and geomagnetic field line.} distribution centered around 90 deg and drift along geomagnetic field lines belonging to the same McIlwain $L$-shell\footnote{An $L$-shell is the surface formed by azimuthal rotation of a dipole field line. In a dipole, $L$ is the radius where a field line crosses the equator; in case of the Earth dipole, it is measured in units of Earth radii. McIlwain's coordinates \citep{McIlwain}, $L$ and $B$ (the magnetic field strength), are pairs describing how far away from the equator a point is located along a given magnetic line at the distance $L$ from the Earth.} where they were produced. Due to magnetospheric transport processes, the antiproton population is expected to be distributed over a wide range of radial distances.

According to the so-called CRAND (Cosmic Ray Albedo Neutron Decay) process \citep{waltcrand,protcrand}, a small fraction of neutrons escapes the atmosphere and decays within the magnetosphere into protons, which become trapped if they are generated with a suitable pitch angle. Such a mechanism is expected to produce antineutrons (through pair production reactions such as $pp\rightarrow ppn\bar{n}$) which subsequently decay to produce antiprotons (CRANbarD). This source is expected to provide the main contribution to the energy spectrum of stably trapped antiprotons and the resulting flux is predicted to be up to several orders of magnitude higher than the antiproton flux from direct $p\bar{p}$ pair production in the exosphere \citep{Fuki,Selesnick}.

The magnetospheric antiproton flux is expected to exceed significantly the galactic CR antiproton flux at energies below a few
GeV. However, predictions differ and suffer from large uncertainties, especially regarding contributions from the CRANbarD
process. A measurement by the Maria-2 instrument \citep{MARIA2} on board the ``Salyut-7'' and ``MIR'' orbital stations
allowed an upper limit on the trapped antiproton-to-proton ratio of $5\cdot 10^{-3}$ to be established below 150 MeV. This Letter
describes the first detection of antiprotons trapped in the inner radiation belt, using the PAMELA satellite-borne experiment.

\section{The PAMELA experiment}
PAMELA was launched from the Baikonur Cosmodrome on 2006 June 15 on board the ``Resurs-DK1'' satellite. The instrument was designed to accurately measure the spectra of charged particles (including light nuclei) in the cosmic radiation, over an energy interval ranging from tens of MeV to several hundred GeV. In particular, PAMELA is optimized to identify the small component of CR antiparticles. Since launch, PAMELA has collected an unprecedented number of antiprotons and positrons, as reported in recent publications \citep{PRL,PRL_BIS,POSFRAC,POSFRAC2}.

PAMELA is built around a permanent magnet spectrometer equipped with a tracking system consisting of six double-sided
micro-strip silicon sensors, which allows the determination of the particle charge and rigidity (momentum/charge) with high
precision. A sampling electromagnetic calorimeter, composed of 44 silicon planes interleaved with 22 plates of tungsten
absorber, is mounted below the spectrometer. A time-of-flight (ToF) system, made of three double layers of plastic scintillator
strips, allows velocity and energy loss measurements, and provides the main trigger for the experiment. Particles leaving
the PAMELA acceptance due to scattering or interactions are rejected by the anticoincidence system. A further scintillator
plane and a neutron detector are placed below the calorimeter, in order to provide additional information about the shower
extension and to improve lepton/hadron discrimination. A detailed description of the PAMELA apparatus along with an
overview of the entire mission can be found elsewhere \citep{PAMELA}.

The satellite orbit (70$^{\circ}$ inclination and 350--610 km altitude) allows PAMELA to perform a very detailed measurement of the
cosmic radiation in different regions of Earth's magnetosphere, providing information about the nature and energy spectra of
sub-cutoff particles \citep{QUASITRAPPED}. The satellite orbit passes through the South Atlantic Anomaly (SAA), allowing the
study of geomagnetically trapped particles in the inner radiation belt.

\section{Antiproton identification}
A clean sample of antiprotons was identified using information combined from several PAMELA subdetectors. Antiprotons are measured in the presence of a considerably larger flux of protons. It is therefore important that particle trajectories are well reconstructed by the tracking system, allowing reliable charge sign separation and a precise estimate of rigidity \citep{PRL}. Strict conditions were placed on the number of position measurements along a track and on the $\chi^{2}$ associated with the track fit procedure, in order to reject protons which were wrongly reconstructed as negatively charged particles due to scattering and to minimize uncertainties on the rigidity measurement.

Selections based on the interaction topology in the calorimeter allow antiproton/electron discrimination. Antiprotons in the
selected energy range are likely to annihilate inside the calorimeter, thus leaving a clear signature. The longitudinal and transverse
segmentation of the calorimeter is exploited to allow the shower development to be characterized. These selections are
combined with $dE/dx$ measurements from individual strips in the silicon detector planes to allow electromagnetic showers to
be identified with very high accuracy.

The particle velocity measurement provided by the ToF and the ionization losses in both the tracker and the ToF planes were used to discard electrons and secondary particles, mostly $\pi^{-}$, produced by CRs interacting in the 2 mm thick aluminum pressurized container in which PAMELA is housed or at the support structures in the upper parts of the apparatus. Further rejection was provided by requiring no activity in the anticoincidence systems and exploiting the ToF and the tracking systems segmentation: in particular, an upper limit was applied on the number of hits close to the reconstructed track, in the two top ToF scintillators and in the tracker planes. The residual contamination was estimated with simulations to be negligible below 1 GV, while it is about 10\% in the rigidity range 1--3 GV \citep{PRL,BRUNO}.

Measured antiproton distributions were corrected by means of simulations to take into account losses due to ionization and
multiple scattering inside the apparatus and, mainly, due to inelastic interactions (annihilation) in the dome. The correction
factor decreases with increasing energy, ranging from 14\% to 9\%. Selection efficiencies were determined using flight data,
which naturally include detector performances. Test beam and simulation data were used to support and cross-check these measurements.
The total systematic error on the measured spectrum includes uncertainties on efficiency estimation, gathering power, livetime, contamination, ionization, and interaction losses. Additional details of the analysis can be found in \citet{PRL_BIS}.

\section{Instrument response}
The factor of proportionality between the antiproton flux and the number of detected antiproton candidates, corrected
for selection efficiencies and acquisition time, is by definition the gathering power of the apparatus. This quantity depends
both on the angular distribution of the flux and the detector geometry. In presence of an isotropic particle flux, the gathering
power depends only on the detector design, and it is usually called the geometrical factor. For the PAMELA apparatus, this
factor depends also on particle rigidity, due to the influence of the spectrometer on particle trajectories.

Fluxes in radiation belts present significant anisotropy since particles gyrate around field lines while moving along them, bouncing back and forth between mirror points. This results in a well-defined pitch-angle distribution. A dependence on the local magnetic azimuthal angle is observed as consequence of the east-west effect. Positively (negatively) charged particles arriving from the east (west) originate from guiding centers located at lower altitudes than PAMELA and thus their flux is significantly reduced by the atmospheric absorption, while the opposite is valid for particles from western (eastern) directions. The resulting asymmetry is more evident for higher rigidity particles since it scales with the particle gyroradius which ranges from $\sim$ 50 km for a 60 MeV (anti)proton, up to $\sim$ 250 km for a 750 MeV (anti)proton.

The flux angular distributions, needed for the estimate of the apparatus gathering power, were evaluated using a trapped
antiproton model \citep{Selesnick}. The calculation was performed using simulations according to the method described in \citet{Sullivan}. The dependency of the directional response function on the satellite orbital position and on its orientation
relative to the geomagnetic field was taken into account. Pitch angle distributions were evaluated at more than 300 points
along the orbit of the Resurs-DK1 satellite in the SAA region and for most probable orientations of PAMELA relative to the
magnetic field lines. The geomagnetic field was estimated on an event-by-event basis using the IGRF-10 model \citep{IAGA}. A mean gathering power, averaged over all PAMELA orbital positions and orientations, was derived. The dependence of the instrument response on particle rigidity was studied by estimating the gathering power at 10 rigidity values in the range of interest. The apparatus gathering power was calculated to be significantly reduced with respect to the geometric factor ($<$3\%), ranging from $\sim0.5$ cm$^{2}$sr at 60 MeV to $\sim10^{-2}$ cm$^{2}$sr at 750 MeV.

\begin{table}[t]
\center
\small
\begin{tabular}{ c c c c c}
\hline
\textbf{Rigidity} & \textbf{Mean kinetic} & \textbf{Observed} & \textbf{$\bar{p}$ flux} & \textbf{$\bar{p}/p$ ratio}\\
\textbf{range (GV)} & \textbf{energy (GeV)} & \textbf{$\bar{p}$ events} & \textbf{(m$^{-2}$s$^{-1}$sr$^{-1}$GeV$^{-1}$)} & (\textbf{$\times10^{-5}$})\\
\hline
\hline
0.35-0.46 & 0.08 & 3 & 8.8 $_{-5.5}^{+6.7}$ $\pm0.9$ & 0.25 $_{-0.16}^{+0.19}$ $\pm0.01$ \\
0.46-0.61 & 0.14 & 9 & 15.3 $_{-4.5}^{+6.4}$ $\pm1.6$ &  0.76 $_{-0.22}^{+0.32}$ $\pm0.01$ \\
0.61-0.81 & 0.23 & 9 & 22.3 $_{-6.6}^{+9.4}$ $\pm2.4$ & 1.44 $_{-0.42}^{+0.61}$ $\pm0.01$ \\
0.81-1.07 & 0.38 & 5 & 43 $_{-19}^{+24}$ $\pm5$ &       6.3 $_{-2.8}^{+3.5}$ $\pm0.4$ \\
1.07-1.41 & 0.60 & 2 & 31 $_{-20}^{+35}$ $^{+4}_{-5}$ & 10.1 $_{-6.3}^{+11.3}$ $^{+0.7}_{-1.1}$ \\
\hline
\end{tabular}
\caption{Summary of antiproton results in the SAA region. The first and second errors represent the statistical and systematic uncertainties, respectively.} \label{table1}
\end{table}

\section{Results}
During about 850 days of data acquisition (from 2006 July to 2008 December), 28 trapped antiprotons were identified
within the kinetic energy range 60--750 MeV. Events with geomagnetic McIlwain coordinates \citep{McIlwain} in the range $1.1 < L < 1.3$ and $B < 0.216$ G were selected, corresponding to the SAA. The fractional livetime spent by PAMELA in this region amounts to the 1.7\% ($\sim4.6\cdot10^{9}$ s).

The propagation of each antiproton candidate was checked using simulation tools which allowed particle trajectories to be traced through the Earth's magnetosphere \citep{TRAJMAP,8bi}. All the identified antiprotons, characterized by a pitch angle near 90 deg, were found to spiral around field lines, bounce between mirror points, and also perform a slow longitudinal drift around the Earth, for a total path length amounting to several Earth radii.

\begin{figure}
\centering
\includegraphics[width=4.0in]{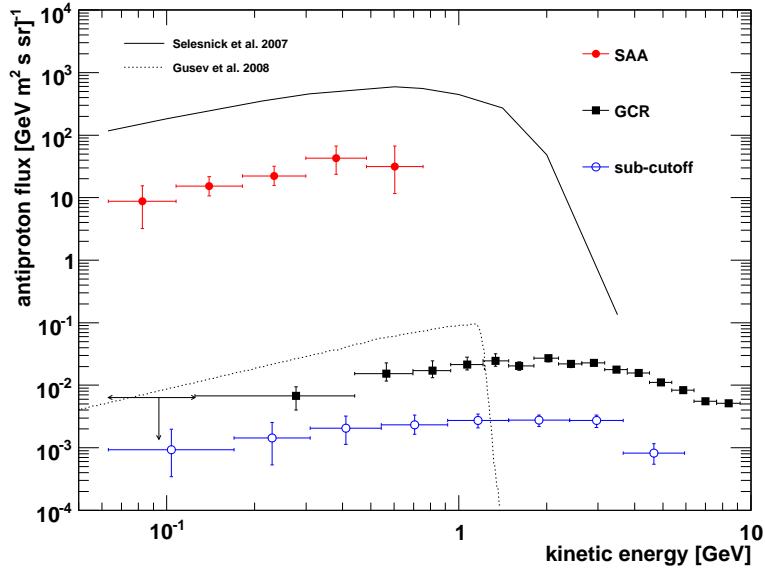}
\caption[h]{The geomagnetically trapped antiproton spectrum measured by PAMELA in the SAA region (red full circles). The error bars indicate statistical uncertainties. Trapped antiproton predictions by \citet{Selesnick} for the PAMELA satellite orbit (solid line),
and by \citet{Gusev} at $L=1.2$ (dotted line), are also reported. For comparison, the mean atmospheric under-cutoff antiproton spectrum outside SAA region (blue open circles) and the galactic CR antiproton spectrum (black squares) measured by PAMELA \citep{PRL_BIS} are also shown.}
\label{fig1}
\end{figure}

\begin{figure}
\centering
\includegraphics[width=4.0in]{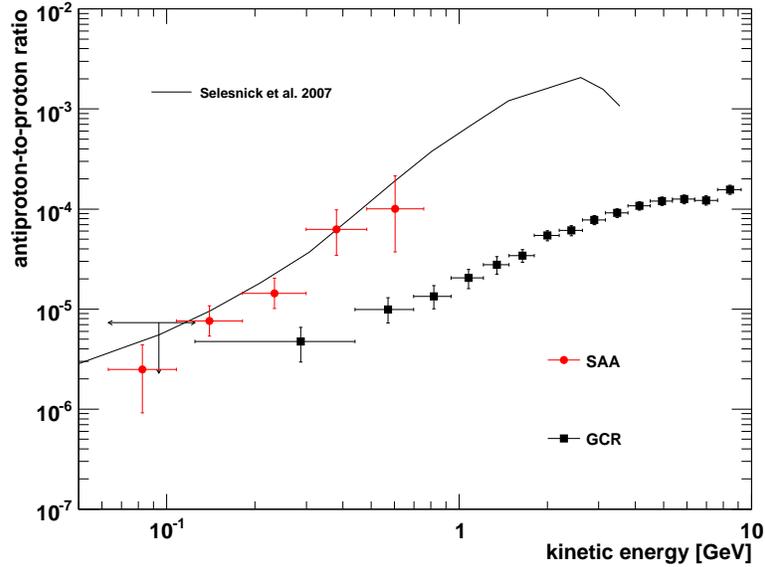}
\caption[h]{The trapped $\bar{p}/p$ ratio measured by PAMELA in the SAA region (red points). Results are compared with predictions by \citet{Selesnick}, denoted by the solid line. The galactic CR $\bar{p}/p$ ratio data \citep{PRL_BIS} are also reported (black squares).}
\label{fig2}
\end{figure}

The spectrum of trapped antiprotons measured by PAMELA in the SAA region is reported in Table \ref{table1} and Figure \ref{fig1}.
Predictions from a CRANbarD model by \citet{Selesnick} for the PAMELA orbit, and an independent calculation by \citet{Gusev} at $L$=1.2, are also shown. Indeed, the estimated magnetospheric antiproton flux is compared with the galactic CR antiproton spectrum \citep{PRL_BIS}, and with the mean spectrum of sub-cutoff antiprotons measured by PAMELA outside the radiation belts ($B>0.23$ G). The
latter result was obtained by selecting particles with a rigidity value lower than 0.8 times the corresponding St\"{o}rmer vertical cutoff\footnote{The vertical cutoff rigidity $R_{VC}$ is the lowest rigidity for which a particle arriving from the zenith direction can access to a given location within the geomagnetic field. It is estimated on an event-by-event basis using orbital parameters according to the St\"{o}rmer formalism: $R_{VC}=14.9/L^{2}$.}. Furthermore, a nearly isotropic flux distribution was assumed. The measured SAA-trapped antiproton flux exceeds the sub-cutoff flux detected outside radiation belts and the galactic CR antiproton flux at the current solar minimum (negative phase $A^{-}$), by four and three orders of magnitude, respectively.

The trapped antiproton-to-proton ratio measured in the SAA is shown in Figure \ref{fig2}, where it is compared with theoretical predictions by \citet{Selesnick} and the antiproton-to-proton ratio measured by PAMELA for galactic particles. The trend reflects the high energy threshold of the pair production mechanism, with values similar to that of the albedo antineutron-to-neutron ratio \citep{Selesnick}.

\section{Conclusions}
Antiprotons trapped in Earth's inner radiation belt have been observed for the first time by the PAMELA satellite-borne experiment.
The antiparticle population originates from CR interactions in the upper atmosphere and subsequent trapping in
the magnetosphere. PAMELA data confirm the existence of a significant antiproton flux in the SAA below $\sim$ 1 GeV in kinetic
energy. The flux exceeds the galactic CR antiproton flux by three orders of magnitude at the current solar minimum, thereby constituting the most abundant antiproton source near the Earth. A measurement of the sub-cutoff antiproton spectrum outside
the SAA region is also reported. PAMELA results allow CR transport models to be tested in the terrestrial atmosphere and
significantly constrain predictions from trapped antiproton models, reducing uncertainties concerning the antiproton production
spectrum in Earth's magnetosphere.

\section*{Acknowledgements}
We gratefully thank Prof. J. Clem for his assistance and support in adapting the ``TrajMap'' simulation program \citep{TRAJMAP} and Prof. R. Selesnick for helpful discussions. We acknowledge support from The Italian Space Agency (ASI), Deutsches Zentrum f$\ddot{u}$r Luftund Raumfahrt (DLR), The Swedish National Space Board, The Swedish Research Council, The Russian Space Agency (Roscosmos) and The Russian Foundation for Basic Research.

\end{document}